\documentclass[amsmath,amssymb,aps,pra,twocolumn,superscriptaddress]{revtex4-2}
\usepackage{graphicx} 
\usepackage{amsmath}
\usepackage{mathtools}
\usepackage{physics}
\usepackage{dsfont}
\usepackage{setspace}
\usepackage{xcolor}
\usepackage{graphicx}
\usepackage{svg}
\usepackage{subcaption}
\usepackage[percent]{overpic}
\usepackage{textgreek}
\usepackage{amssymb}
\usepackage{comment}
\usepackage{tabularx}

\usepackage[justification=justified, singlelinecheck=false]{caption}
\usepackage[labelformat=simple,justification=justified]{subcaption}
\usepackage{float}
\usepackage{hyperref}
\usepackage{titlesec}
\newcommand{\kb}[1]{| #1 )}

\renewcommand{\raggedright}{\leftskip=0pt \rightskip=0pt plus 0fil}
\begin{document}

\title{Entanglement generation from degeneracy breaking of superradiant Liouvillian steady states by local operations}

\author{Adam Burgess}%
\affiliation{SUPA, Institute of Photonics and Quantum Sciences, Heriot-Watt University, Edinburgh, EH14 4AS, UK}

\author{Erik M.~Gauger}%
\affiliation{SUPA, Institute of Photonics and Quantum Sciences, Heriot-Watt University, Edinburgh, EH14 4AS, UK}

\begin{abstract}
Superradiance is a collective optical effect where emitters cooperatively radiate with a maximal intensity scaling proportional to the square of the number of emitters. 
Despite the superradiant cascade transitioning through highly entangled states, it has been shown that, for dynamics along the Dicke ladder, superradiance need not generate entanglement between the emitters. By contrast, we show how introducing incoherent local processes -- normally considered detrimental to both entanglement and superradiance -- can generate high levels of inter-emitter entanglement. We develop a general framework, using degenerate non-Hermitian perturbation theory, for the study of superradiant Liouvillian steady-state degeneracy breaking by these local operations from a few emitters up into the thermodynamic limit. From this, we are able to show that entanglement becomes ubiquitous in this regime and demonstrate its robustness against other forms of symmetric environmental noise. In the regime where pumping dominates, we find that population accumulates towards the maximally entangled minimal pseudo-spin states, corresponding to multi-emitter singlet states. This provides a deeper insight into many-body entanglement and the curious connection between local operations and entanglement generation. 
\end{abstract}
\maketitle
    The Dicke model is a foundational framework in quantum optics that describes the collective coupling of an ensemble of ($N$) two-level atoms to a common electromagnetic mode~\cite{Dicke1954,Garraway2011,Kirton2019,Holzinger2026,lyne2026dickesuperradiancedegeneratequantum,Pannier2026,holzinger2026symbolicquantumtrajectorymethodmultichannel,Pannier2025}. Coupling to a single mode leads to the formation of polaritonic states and coherent collective Rabi oscillations~\cite{scully1997quantum,AllenBook}. The extension to many modes exhibits a wide range of dynamical behaviour, determined by the structure of the photonic environment. This includes Markovian decay in free space~\cite{TheoryOQSBook}, damped oscillations in lossy cavities~\cite{Garraway1997,Burgess2023,Gonzalez2026fromsuperradianceto}, and excitation trapping in photonic band-gap systems~\cite{Garraway1997,Burgess2022}. An emblematic prediction of the Dicke model is superradiance: a burst of radiation with intensity scaling as $N^2$, arising from emission from highly entangled states with `greater-than-the-sum-of-their-parts' dipole strength, along with the time-reversed process of superabsorption~\cite{Burgess2025,Quach2022}.   Beyond intensity, another signature of superradiant emitters is photon bunching in Hanbury Brown-Twiss experiments, where cooperativity leads to anti-dips in photon coincidences~\cite{Kim2018,Grim2019,Cygorek2023,Banerjee2025}. 
While semiclassical approaches successfully reproduce key features of Dicke dynamics, including the burst-time scaling ($t_B \propto \ln N$)~\cite{MacGillivray1976,Arecchi1972}, the phenomenology~\cite{Holzinger2026,lyne2026dickesuperradiancedegeneratequantum,Pannier2026,holzinger2026symbolicquantumtrajectorymethodmultichannel,Pannier2025,Wellnitz2020,Sandik2025,Feist2018} and the contribution of quantum correlations to the dynamics remains an active area of investigation~\cite{Rosario2025,Wolfe2014,Yu2016,Gonzalez2026fromsuperradianceto}. 

Within an open quantum systems framework, the many-body state of the emitters undergoing superradiance evolves through Dicke states, where all states, besides the fully inverted and ground states, are entangled. However, determining whether the resulting mixed state is separable poses a considerable challenge~\cite{Toth2010,Wolfe2014}.
Recent insights have shown that superradiant decay along the maximal Dicke ladder (pseudo-spin $j=N/2$) from the fully inverted state does not generate entanglement between emitters due to a possible decomposition into non-entangled coherent pseudo-spin states~\cite{Rosario2025}. Conversely, non-fully-inverted initial states or accounting for dynamics on minor Dicke ladders ($j<N/2$) can generate entanglement~\cite{Tacchino2018}, indicating that the connection between superradiance and entanglement is manifestly multi-faceted and subtle. 

\begin{figure}[t!]
    \centering
    \begin{subfigure}[t]{\linewidth}
        \centering
        \includegraphics[width=0.9\linewidth]{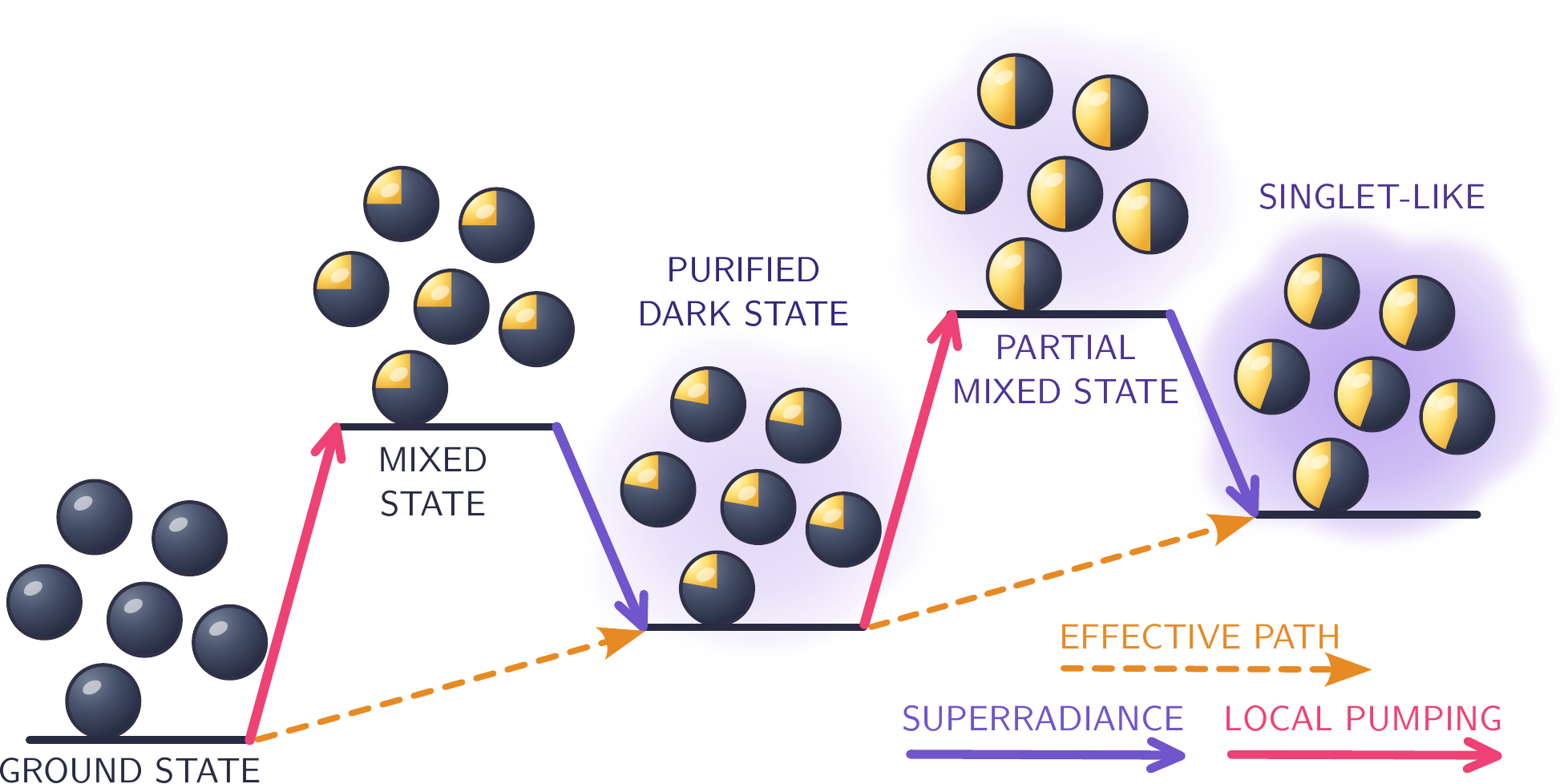}
    \end{subfigure}%
    \vspace{5pt}
    
    \begin{subfigure}[b]{0.7\linewidth}
        \centering
        \includegraphics[width=\linewidth]{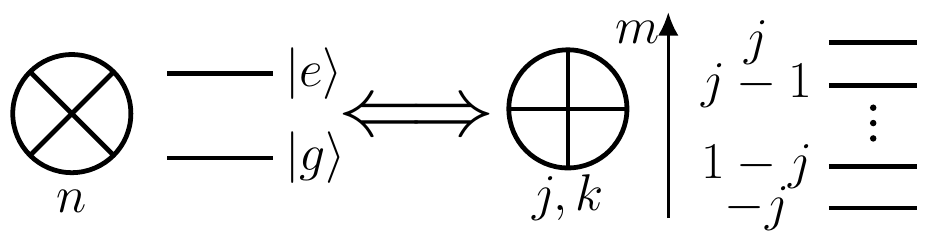}
    \end{subfigure}%
    \vspace{5pt}
    
     \begin{subfigure}[b]{0.4\linewidth}
     \centering
\begin{overpic}[width=\linewidth]{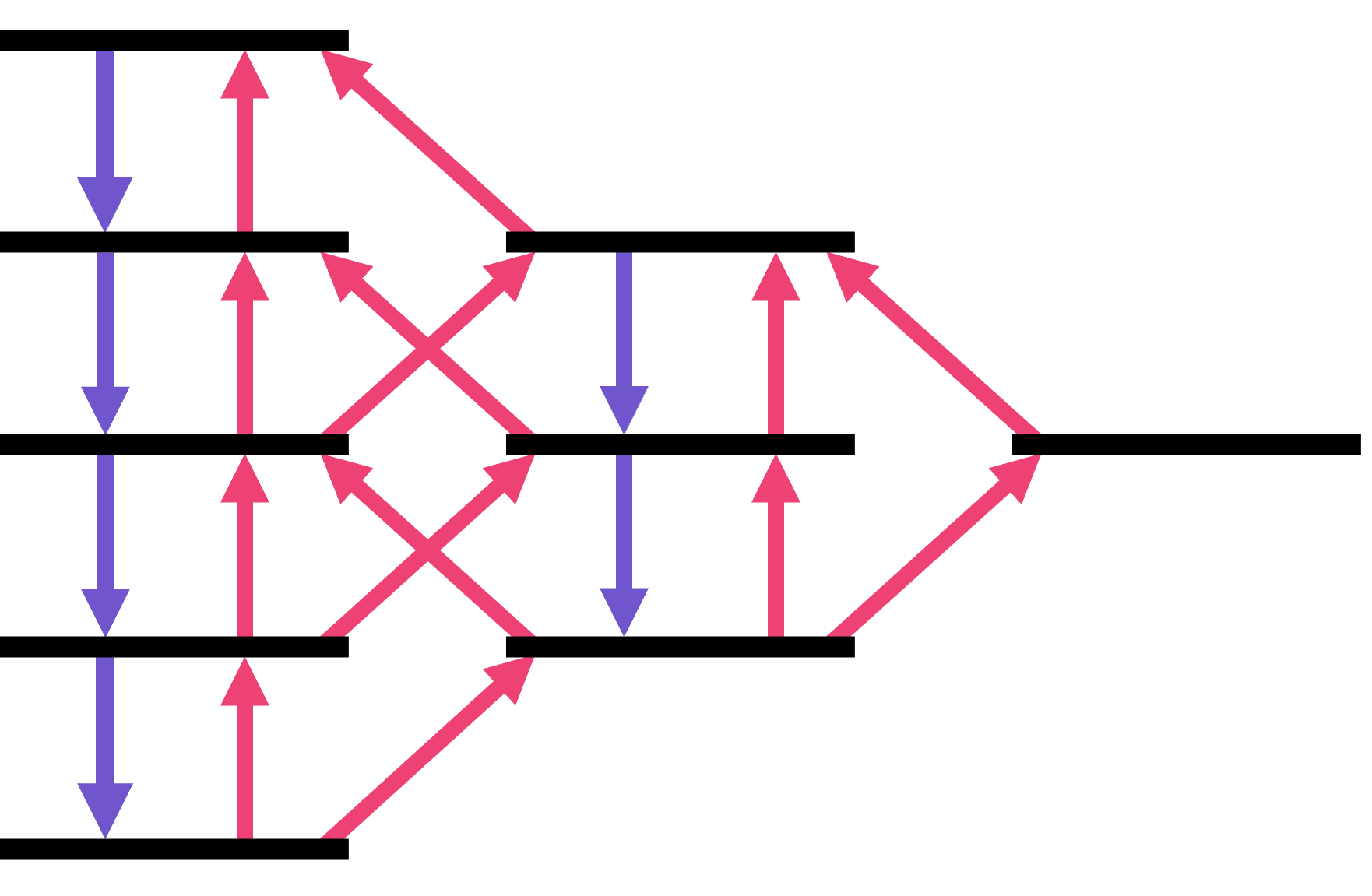}
    \put (-5,225) {\textbf{(a)}}
    \put (-5,110) {\textbf{(b)}}
     \put (-5,70) {\textbf{(c)}}
     \put (100,45) {$\underrightarrow{\gamma_P\ll\gamma_o}$}
 
    \end{overpic}
    \end{subfigure}
    ~\hspace{33pt}
     \begin{subfigure}[b]{0.4\linewidth}
        \centering
        \includegraphics[width=\linewidth]{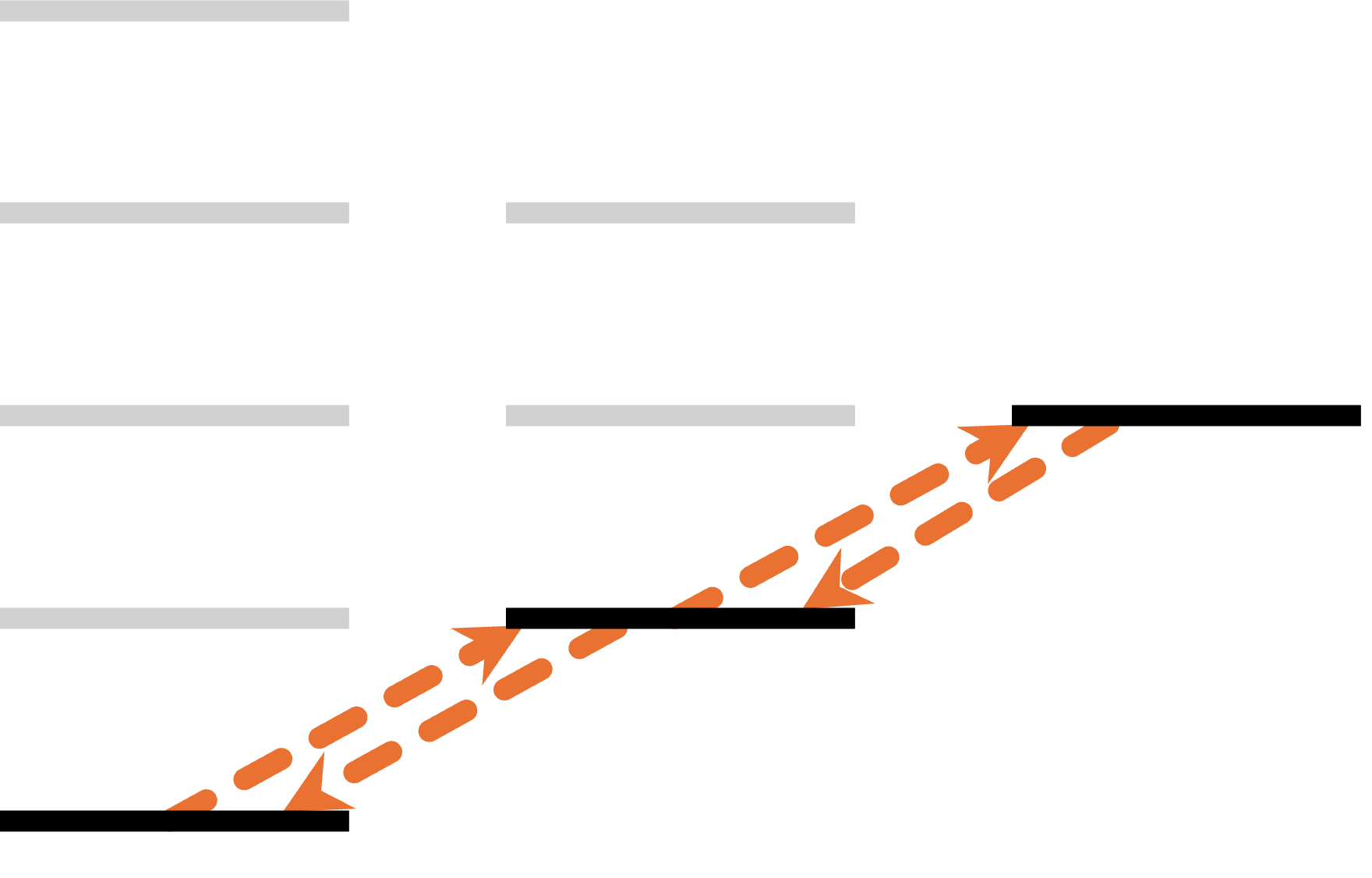}
    \end{subfigure}%
    
    \caption{(a) A graphic depiction of the process by which emitters are ratcheted up the superradiant dark manifolds by weak local pumping towards multi-emitter singlet states. The circles depict emitters and the proportion of yellow filling, their average population. After pumping, optically bright states decay, leaving entangled dark states. The purple cloud depicts a degree of entanglement between the emitters. (b) Depiction of the equivalence between the tensor product space of $N$ two-level systems and the tensor sum space of Dicke ladders with different total pseudo-spin $j$ and multiplicity $k$. (c) Level diagram for the transitions in the pseudo-spin basis due to superradiant decay (purple arrows) and pumping (red arrows) for four emitters. 
    Weak pumping breaks the degeneracy of the superradiant steady states, enabling 
    ratcheting up the dark states.}
       \label{fig:fig1}
\end{figure}

In open, driven realizations of superradiant systems -- such as ultra-cold atoms in optical cavities ~\cite{Baumann2010,Baumann2011}, where the superradiant phase transition has been observed experimentally -- dissipation plays a crucial role. Photon loss preserves the total pseudo-spin and is well understood~\cite{Emary2003,Kirton2017}. In contrast, incorporating individual dephasing and dissipation leads to significantly richer and more complex behavior~\cite{DallaTorre2016,Kirton2017,Shankar2021}. Due to the exponential scaling of the Hilbert space in such many-body systems, permutation-invariant frameworks~\cite{Chase2008,Richter2015,Kirton2017,KolmerForbes24} were developed to greatly reduce the effective Hilbert space dimension by exploiting weak permutation symmetry~\cite{Albert2014}. Utilizing this method, it was shown that even infinitesimal dephasing suppresses the superradiant phase transition. However, including pseudo-spin relaxation restores the transition~\cite{Kirton2017}. This further suggests that non-classical features and additional dissipation channels play an important role in the behavior of superradiant systems. 

In this Letter, we develop a framework based on degenerate non-Hermitian perturbation theory that leverages the permutation symmetry of the local operations associated with dephasing, pumping, and independent decay. We use this framework to identify and characterize the breaking of the degeneracy of the superradiant Liouvillian steady states. This yields a unique Liouvillian degeneracy-broken steady state (LDBSS); by degeneracy, we refer to the eigenvalues of the non-Hermitian Liouvillian, not the system Hamiltonian. We determine the entanglement present within these LDBSSs using the negativity entanglement monotone. We find that, when pumping dominates the local processes, the LDBSSs can have a large degree of entanglement, even in the thermodynamic limit. This can be understood as the local pumping ratcheting excitations up the optically dark states towards the multi-emitter singlet states. A depiction of this is given in Fig.~\ref{fig:fig1}(a).

Beginning with the multi-mode Dicke model Hamiltonian, written as 
\begin{equation}
    H = \sum_{i=1}^N\omega_0\sigma_+^i\sigma_-^i + \sum_{i,\lambda} g_{i\lambda} (a_{\lambda}+a^\dag_\lambda)\sigma^i_x+\sum_\lambda \omega_\lambda a^\dag_\lambda a_\lambda,
\end{equation}
where $\omega_0$ is the transition energy of each emitter, $\sigma_+^i = \ket{e_i}\bra{g_i} = (\sigma_-^i)^\dag$ is the raising operator of emitter $i$, $\sigma_x^i = \sigma_+^i+\sigma_-^i$, $g_{i\lambda}$ is the coupling strength of the $i^\text{th}$-emitter to the $\lambda$ photon mode which has creation (annihilation) operator $a_\lambda^{(\dag)}$ and energy $\omega_\lambda$. 
For free-space modes coupled to identical emitters with deep sub-wavelength separations, and within the secular-perturbative regime, tracing over the photon modes yields the quantum-optical master equation for the emitters in Lindblad form~\cite{TheoryOQSBook}.
The Lindbladian associated with a jump operator $L$ and decay rate $\gamma_L$ is 
$
    \mathcal{L}(\gamma_L,L)\rho =\gamma_L\left(L\rho L^\dagger -\frac{1}{2} \left\{L^\dagger L, \rho\right\} \right)
$,
where $\rho$ is the reduced density matrix for the emitter system. 
For $N$ superradiant emitters, the quantum optical master equation has a jump operator given by the collective decay operator $\sigma_-^S=\sum_{i=1}^N\sigma_-^i$, i.e.~the superradiant Lindbladian takes the form  $
    \mathcal{L}_{o} =  \mathcal{L}\{\gamma_{o},\sigma_-^S\} $.
When studying superradiant decay, it is useful to consider the irreducible representation for the emitter system, where we utilise the pseudo-spin states as a basis, i.e. for any state 
$\ket{\psi} = \sum_{\mathbf{v}}c_\mathbf{v}\bigotimes_i \ket{v_i}=\sum_{j,m,k}c_{jmk}\ket{j,m,k},
$
where $\ket{v_i}$ are states in the `local' emitter basis and $\ket{j,m,k}$ corresponds to states with total pseudo-spin $j$, $z$-projection of the pseudo-spin $m$, such that $\sum_i\frac{1}{2}\sigma_z^i\ket{j,m,k}=m\ket{j,m,k}$, and multiplicity parameter $k$. The multiplicity plays an important role as it accounts for the multitude of dark states present in many-emitter systems. The pseudo-spin ($c_{jmk}$) and local ($c_\mathbf{v}$) basis state amplitudes are related via Clebsch-Gordan coefficients~\cite{cohentannoudji1977vol2}. A graphic depiction of this relationship between the local basis of the emitters and the collective pseudo-spin basis is given in Fig.~\ref{fig:fig1}(b). 

In the following, we consider Lindbladians that are symmetric under the exchange of emitters. As superradiance acts on collective states and conserves the total pseudo-spin, it is useful to introduce the permutation-symmetric pseudo-spin (PSPS) states, defined as mixed states over the pseudo-spin multiplicities with $\overline{\ket{j,m}\bra{j,m'}}=\sum_k\ket{j,m,k}\bra{j,m',k}/d_j^N$
where $d_j^N$ is the multiplicity of the $j$ manifold, and the bar denotes a mixed state over equivalent pseudo-spin ensembles~\cite{Chase2008}. For convenience, we adopt the notation $\kb{j,m}=\overline{\ket{j,m}\bra{j,m}}$. This representation is natural for superradiant decay as the decay operator restricts states to fixed $j$ manifolds as $\sigma_-^S\ket{j,m,k}=s_{j,m}\ket{j,m-1,k}$. It follows that the superradiant Lindbladian maps PSPS states as 
$
     \mathcal{L}_{o} \kb{j,m} =\gamma_{o} s_{j,m}^2(\kb{j,m-1}-\kb{j,m})
$
with $s_{jm}^2 = j(j+1)-m(m-1)$. As the superradiant decay Lindbladian maps states to lower $z$-projection states, there is a degenerate manifold of steady states for all the lowest $z$-projection states for each $j$ manifold, that is $\text{ker}\{\mathcal{L}_{o}\} \supset \text{span}\{\kb{j,-j}\}$.

In realistic emitter systems, superradiant behaviour will be accompanied by other dynamical processes. We consider three representative cases here. Each takes the form of a symmetric local Lindbladian (SLL) that preserves the PSPS state structure. These SLLs take the form $\mathcal{L}_{SL}(\gamma_s,\sigma_s)=\sum_{i=1}^N\mathcal{L}(\gamma_s,\sigma_{s}^i)$ for $s=\{+,-,z\}$. We refer to these as local, as they act independently on each emitter. Such SLLs map PSPS states to PSPS states as 
\begin{align}
    \mathcal{L}_{SL}(\gamma_s,\sigma_s) \kb{j,m} &= \gamma_s\big(\eta_{s}^{j,m}\kb{j,m}+\tau_{s}^{j,m}\kb{j,m_s}\\&+ \beta_s^{j,m}\kb{j-1,m_s} \nonumber+\kappa_s^{j,m}\kb{j+1,m_s}\big),\nonumber
\end{align}
for $s=+,-,z$ where $m_+=m+1$, $m_-=m-1$ and $m_z=m$ and the transition elements $\eta_{s}^{j,m},\tau_{s}^{j,m},\beta_s^{j,m}\kappa_s^{j,m}$ have been calculated previously in Ref.~\cite{Chase2008}.
In many relevant optical experiments, we wish to excite the emitters so subsequent superradiant decay can occur; often, it is practical to perform such excitation via weak continuous-wave driving leading to incoherent pumping~\cite{Hallett2026}. Such a pumping process can be modeled via the SLL
    $\mathcal{L}_P = \mathcal{L}_{SL}(\gamma_P,\sigma_+)$.
This pumping SLL excites emitters individually. From the definition of the SLL, we see that pumping increases the $z$-projection of the PSPS state ($m$) by one and changes the total pseudo-spin ($j$) by at most one. This is depicted in the level scheme of four emitters in Fig.~\ref{fig:fig1}(c). 
The next two SLLs we will consider are associated with environmental noise. For pure dephasing between the emitters, which could manifest from fluctuations in the emitters' geometries or local environment, the noise can be modeled by a dephasing Lindbladian of the form 
$ \mathcal{L}_D = \mathcal{L}_{SL}(\gamma_D,\sigma_z)$~\cite{Leggett1987}.
The final SLL we shall consider is associated with independent decay of emitters. This can occur due to non-radiative decay processes or strong coupling to phonons leading to the partial breakdown of superradiance; this is particularly relevant for solid-state emitters~\cite{Wiercinski2024}. Such an SLL takes the form $\mathcal{L}_{I}=  \mathcal{L}_{SL}(\gamma_I,\sigma_-)$~\cite{KolmerForbes24}. 

As all three SLLs act symmetrically yet independently on the emitters, they introduce decoherence between them. Consequently, when these processes dominate the dynamics, the steady state is expected to be separable. We therefore anticipate that the weak-action regime will provide the clearest insight into entanglement generation.

To study the weak effects of the SLLs, we deploy degenerate non-Hermitian perturbation theory, where we assume $\gamma_{P,D,I}\ll\gamma_{o}$, and that the total Lindbladian is given by
$    \mathcal{L} = \mathcal{L}_{o}+\epsilon(\mathcal{L}_D+\mathcal{L}_{P}+\mathcal{L}_{I})$, where $\epsilon$ is our perturbative perameter.
Expanding $\rho=\rho_0+\epsilon \rho_1 + \mathcal{O}(\epsilon^2)$, we find, to first order in $\epsilon$, the degeneracy of the superradiant Liouvillian steady states is lifted by the consistency equation $ \mathcal{L}_{o}\rho_1=-(\mathcal{L}_D+\mathcal{L}_{P}+\mathcal{L}_{I})\rho_0$.
 By relabeling the PSPS states with states that count the number of excitations  $\kb{n}=\kb{N/2-n,n-N/2}$, an element of the steady state space of the superradiant Lindbladian has the form  $\rho_0 = \sum_n p_n\kb{n}$. Thus, we can solve the consistency equation, and by comparing coefficients of the states, we generate a two-step recursion relation for the state probabilities of the form $p_n = f_{n}p_{n-1}+g_np_{n-2}$
where $f_n$ and $g_n$ are functions of the rates $\gamma_P,\gamma_I,\gamma_D$, explicitly defined in the SM. Recent work, neglecting dephasing, found a similar recursion relation from an ansatz informed by numerical insights that agree with those analytically derived here in the appropriate limits~\cite{shankar2026dynamicalaspectssteadystatesubradiance}. Our procedure is equivalent to adiabatically eliminating non-superradiant steady states and calculating the population transfer between the superradiant steady states due to the SLLs. A depiction of this is shown in Fig.~\ref{fig:fig1}(c).

\begin{figure}
    \centering
    \begin{overpic}[width=\linewidth]{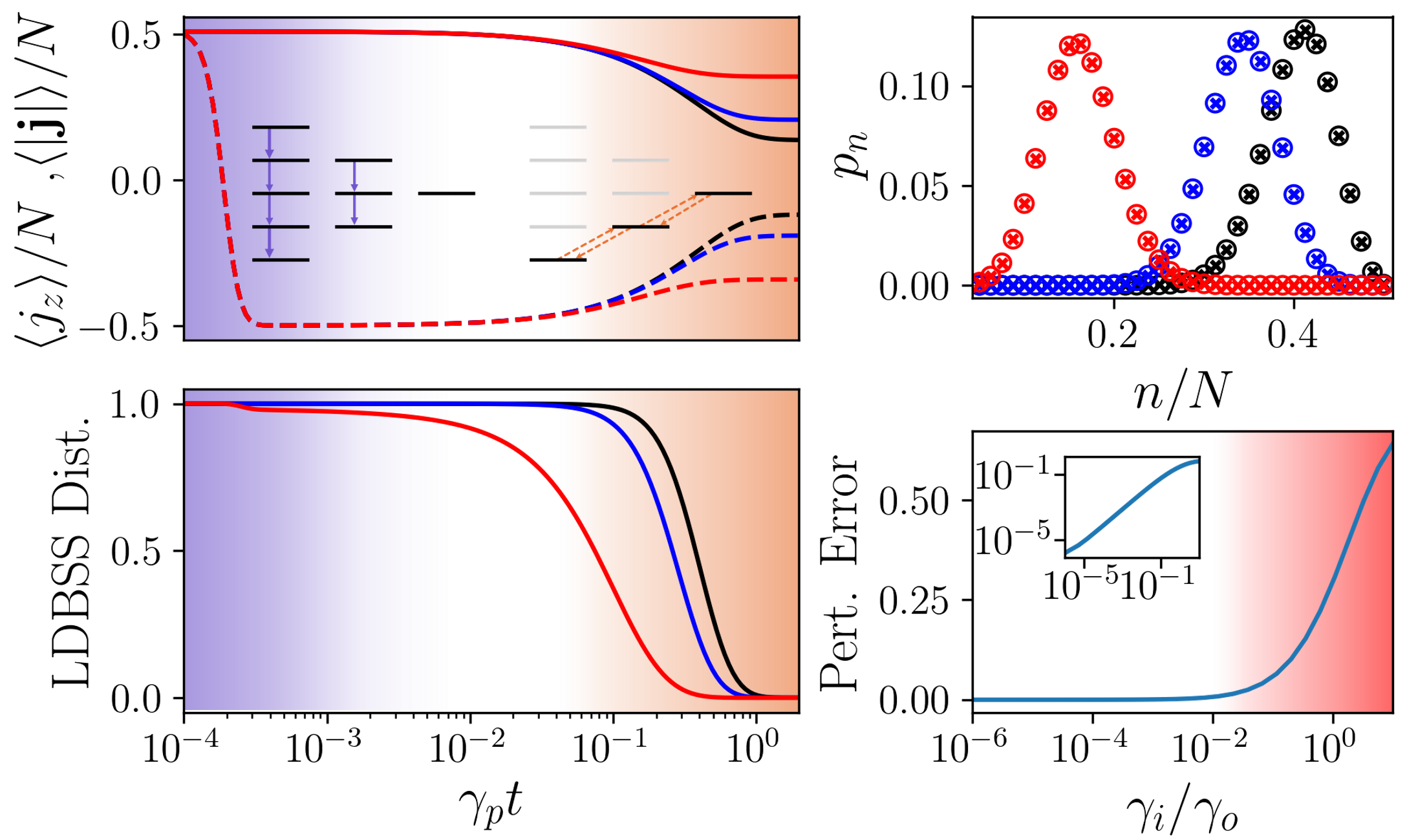}
    \put(-1,60) {\textbf{(a)}}   
    \put(36.5,52) {\footnotesize D.B.}
    \put(17.,52) {\footnotesize S.R.}
    
    \put(-1,30) {\textbf{(b)}}    
    \put(57,60) {\textbf{(c)}}    
    \put(57,30) {\textbf{(d)}} 
    \end{overpic}
    
    \caption{ (a)  The exact time evolution of $\langle j_z/N\rangle$ (dashed) and $\langle|\mathbf{j}|/N\rangle$ (thick) for a 50-emitter system starting in the fully inverted state. The purple (orange) shaded region depicts the regime where superradiance (Liouvillian degeneracy breaking) processes are the dominant dynamical process. (b) The dynamics of a distance measure (infidelity $1-F$) between the exact system and the LDBSS solution. (c) The steady-state probabilities for the symmetric pseudo-spin states of $n$-excitations  $\kb{n} =\kb{N/2-n,n-N/2}$, for $80$ emitters, using the exact permutation-invariant method (crosses) and the recursion relation (circles). The rates $(\gamma_P/\gamma_o,\gamma_D/\gamma_o,\gamma_I/\gamma_o)$ were set to  $(10^{-3},10^{-3},10^{-3})$ (black), $(10^{-3},5\times10^{-3},10^{-3})$ (blue)  and $(10^{-3},10^{-3},5\times10^{-3})$ (red). (d) The perturbation error (infidelity) of the exact PSPS steady state and the LDBSS for 80 emitters as a function of the relative strength of the SLL rates and the optical decay rate; the inset depicts the same plot on a log-log scale showing the linear nature of the error, and the red region depicts the regime where errors exceed $1\%$.}
    \label{fig:Pert_v_Perm}
\end{figure}

This recursion relation can then be solved numerically to yield the LDBSS. From this, we recover the ground state $\rho=\kb0$ when pumping is removed ($\gamma_P=0$), as only operations that reduce or maintain the pseudo-spin $z$-projection are possible. Conversely, when pumping is stronger than dephasing and independent decay ($\gamma_P>\gamma_I,\gamma_D$), excitations ratchet along the dark states of each $j$-manifold towards the minimum total pseudo-spin $j$ states ($j=1/2,0$). As we shall show, these states are among the most entangled in $N$-emitter systems. Fig.~\ref{fig:fig1}(a) provides a graphical depiction of this mechanism. In Fig.~\ref{fig:Pert_v_Perm}(a), we calculate the expectation values $\langle j_z\rangle$ and $\langle |\mathbf{j}| \rangle$, which indicate the rung and ladder, respectively, of the multi-emitter state. We also show in Fig.~\ref{fig:Pert_v_Perm}(b) the evolution of the infidelity $(1-F)$ to the LDBSS, for a 50-emitter system initialized in the fully inverted state and subject to three different strengths of the local processes. At early times, the system undergoes superradiant decay along the maximal Dicke ladder, relaxing to the ground state. It remains there until the perturbative Liouvillian degeneracy-breaking processes become relevant, at which point population begins to leave the maximal Dicke ladder, as evidenced by the decrease in $\langle |\mathbf{j}| \rangle$. The system subsequently relaxes toward the LDBSS obtained from the recurrence relation on a timescale of $\gamma_P^{-1}$.

In Fig.~\ref{fig:Pert_v_Perm}(c), we solve the recurrence relation for an 80-emitter system over a range of perturbative decay rates and compare the resulting LDBSS with the steady state obtained from the exact permutation-invariant method~\cite{Kirton2017,Shammah2018}. We find excellent agreement between the two approaches, with infidelities of order $10^{-4}$, validating the recurrence-relation framework. The anticipated error from this perturbation approach is first order in the decay rates. This behavior is illustrated in Fig.~\ref{fig:Pert_v_Perm}(d), where we plot the infidelity as a function of the ratio between the SLL rates and the superradiant decay rate, $\gamma_i/\gamma_o$. Over a broad range of relative strengths, the infidelity is approximately linear in this ratio, as expected.  The permutation-invariant approach has a density matrix of dimension $\frac{1}{6}(N+3)(N+2)(N+1)$. Conversely, this LDBSS method enables consideration of systems orders of magnitude larger, as the density matrix has dimension $\lfloor N/2\rfloor+1$. However, these results apply only in the perturbative limit and in calculating the steady state. 

\begin{figure}[t]
    \centering
    \begin{overpic}[width=\linewidth]{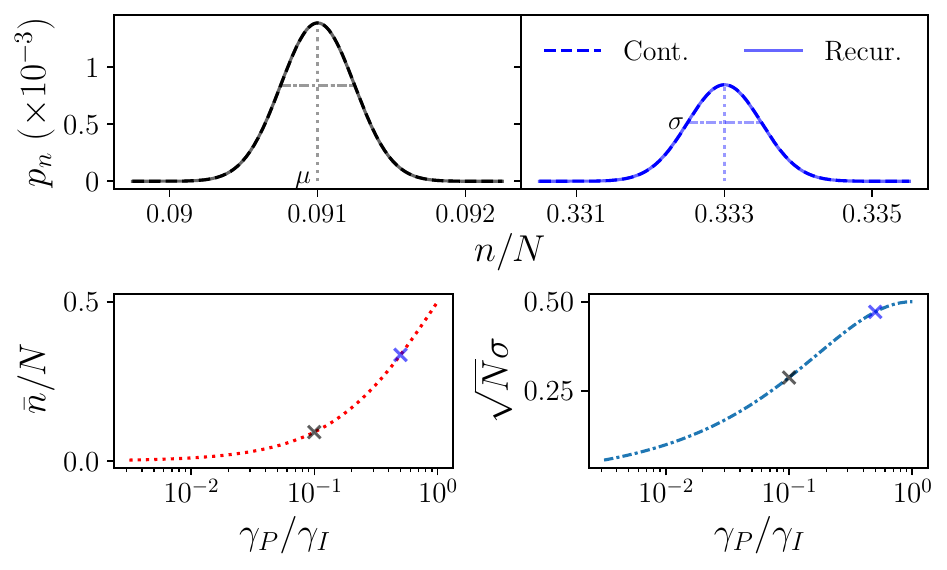}

    \put(-2,60) {\textbf{(a)}}
    \put(-2,27) {\textbf{(b)}}
    \put(49,27) {\textbf{(c)}}
    
    \end{overpic}
    
    \caption{(a) The probabilities of being in the pseudo-spin state with $n$-excitations  $\kb{n} =\kb{N/2-n,n-N/2}$ at the steady state for $10^6$ emitters, calculated numerically using the recursion relation (thick), and using the continuous approximation (dotted); the average difference was $\sim 10^{-8}$. The rates were set to $\gamma_D=2\gamma_P=\gamma_I$ (blue), $\gamma_D=10\gamma_P=\gamma_I$ (black). The (b) average excitation number $\bar{n}$ and (c) standard deviation $\sigma$ of the probability distribution for the emitters as a function of the ratio $\gamma_P/\gamma_I$, containing the values from (a) (crosses).}
    \label{fig:WKB_vs_Pert}
\end{figure}

We now consider the large $N$ regime. This is achieved by considering the quasi-continuous fractional excitation $x=n/N$ and taking the limit $N\rightarrow \infty$. This leads to the coefficients of the recursion relation becoming
$ f(x)  = 1-g(x)$ and $
    g(x) = -{(1-x){r}}/{x}$ with
${r}=\gamma_P/\gamma_I$. Interestingly, we note that in the large $N$ limit, the effects of dephasing are completely suppressed, leaving only the effects of pumping and independent decay. 
Similar dephasing suppression has been seen in polaritonic systems~\cite{Rouse2024}.  Assuming that the probabilities have the following form in the thermodynamic limit $p_n=p(x) \propto e^{NS(x)}$, we can approximate the relative terms in the recurrence relation by considering the Taylor expansion of $S(x)$ to first order as $p_{n-k} \propto e^{NS(x)-kS'(x)}$. 
The recurrence relation can now be converted to a differential equation, which can be solved for $S'(x) =  \ln\{|g(x)|\}$. From this, we can determine that the maximum of the probabilities occurs when $x=\mu={{r}}/({1+{r}})$. Expanding $S(x)$ to second order, we can determine that for $\mu$ away from the boundary $\mu<0.5$, the probability distribution is given by $ p(x)\propto \exp\{{-N\frac{{r}}{2\mu^2}\big(x-\mu\big)^2}\}$. That is, the probabilities of the pseudo-spin states follow a normal distribution with mean $\mu$ and variance $\sigma^2 ={\mu^2}/({N{r}})$. As we increase the number of emitters, this tends to a $\delta$-function at $\mu$. The average number of excitations is $\langle n\rangle = N\mu$, and the variance in the number of excitations is $\Delta n^2=N{\mu^2}/{r}$. In Fig.~\ref{fig:WKB_vs_Pert}(a), we show the solution to the recurrence relation along with the continuous approximation result for $10^6$ emitters for different relative rates, and again we find strong agreement between these approaches with errors below $10^{-8}$. In Fig.~\ref{fig:WKB_vs_Pert}(b,c), we show the mean and variance of the excitations for varying relative strengths $\gamma_P/\gamma_I$. Interestingly, these statistics match those of independent emitters with pumping rate $\gamma_P$ and decay rate $\gamma_I$, despite the dominant feature being the superradiant decay rates. The main distinction here is that, as we will show, these states are entangled, while for independent emitters the density matrix is an unentangled tensor product of emitter states.

 To validate the existence of entanglement in the LDBSSs, we utilize the negativity $\mathcal{N}(\rho)$ entanglement monotone, where $\mathcal{N}(\rho)>0$ implies that a state is entangled~\cite{Horodecki1998,Vidal2002,Plenio2005}. The negativity of a state $\rho$ is defined as $    \mathcal{N}(\rho) = \sum_{\lambda_i<0}{|\lambda_i|}$,
 where $\lambda_i$ are the eigenvalues of the density matrix $\rho^{\perp}_A$ the partial transpose of the state $\rho$ with respect to a bipartition across $A$.  We consider a partition consisting of a single emitter and the remaining $(N-1)$ emitters (i.e. $1:N-1$), a natural choice given the simplicity of the Clebsch–Gordan coefficients and the permutation symmetry of the system. One finds that for the PSPS states that appear in the LDBSS, the negativity of a single state is 
$\mathcal{N}(\kb{j,-j})=\frac{1}{2\sqrt{1+2j}}(1-\frac{2j}{N})$. 
This clearly shows that maximal entanglement occurs at minimal $j$ value, which for an even number of emitters is $j=0$, leading to $\mathcal{N}= 1/2$. One can generalize this result to the mixed states associated with the LDBSS, and we find 
\begin{align}
    &\mathcal{N}(\rho) = \sum_{j} \frac{1}{2(1 + 2 j) N}\bigg[{\bigg({{j^2  (2 + 2 j + N)^2 p_j^2}{}}}+\\&\frac{ (1 + 2 j)^2 (2 - 2 j + 
    N)^2 p_{j-1}^2}{1 - 2 j}\bigg)^{1/2}\nonumber-p_jj(2 + 2j + N)\bigg],
\end{align}
see SM for more details.
This has the interesting feature of always producing a positive value for all LDBSSs, except the ground state ($p_{j=N/2}=1$), showing that for any positive pumping, there is a non-zero entanglement at the steady state. A surprising corollary of this is that any permutation-invariant initial condition with support outside the maximal pseudo-spin manifold $j=N/2$, undergoing only superradiant decay, will have an entangled steady state. This can be seen from 
\begin{equation}
\sum_{j,m,m'}\rho_{j,m,m'}\overline{\ket{j,m}\bra{j,m'}}\xrightarrow{t\rightarrow\infty}\sum_{j}(\sum_m\rho_{j,m,m})\kb{j,-j}.
\end{equation}

\begin{figure}[t]
    \centering
    \begin{subfigure}[t]{\linewidth}
    \begin{overpic}[width=\linewidth]{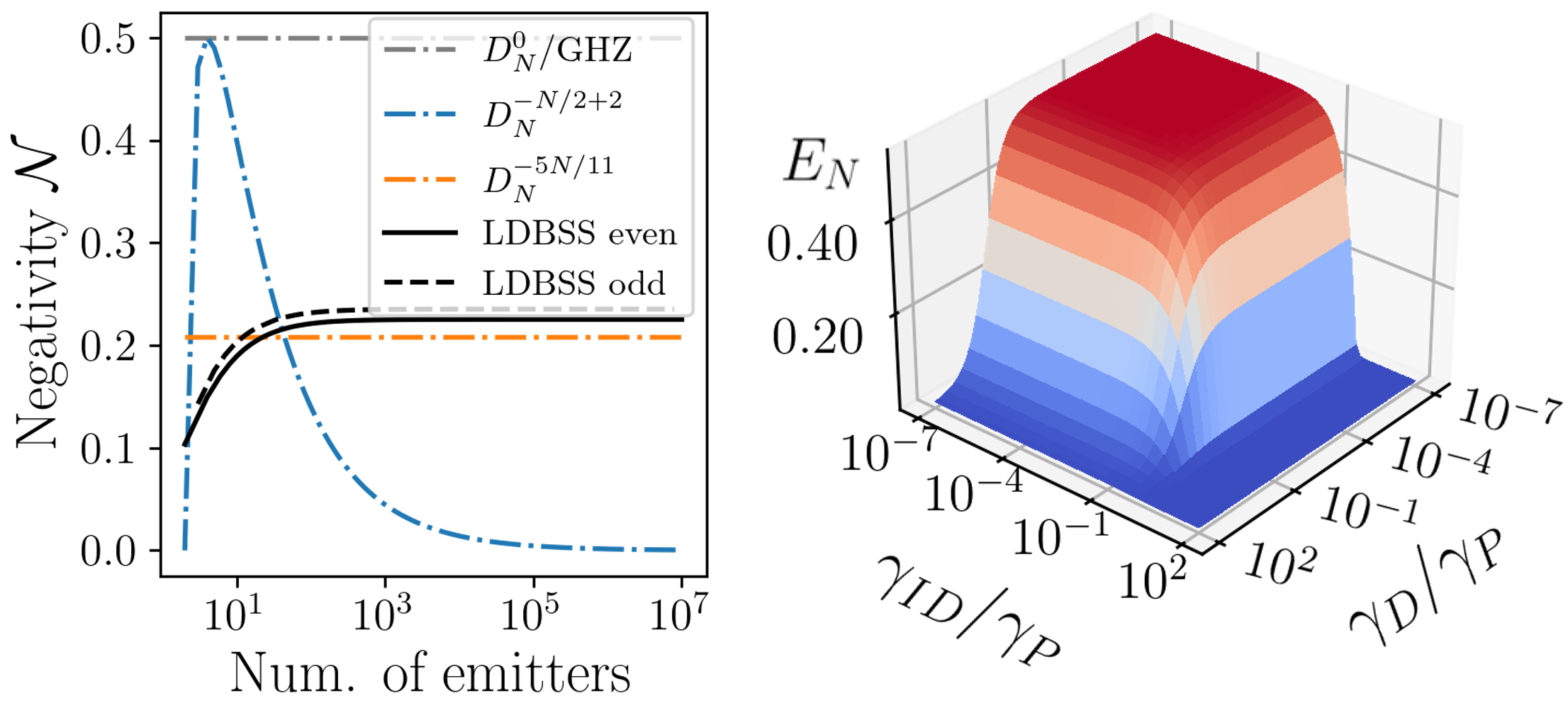}
    \put (-1,44) {\textbf{(a)}}
    \put (46,44) {\textbf{(b)}}
    \put (-1,-3) {\textbf{(c)}}
    \end{overpic}
    \end{subfigure}
    
    \vspace{5pt}
    
    \begin{subfigure}[t]{\linewidth}
    \hspace{30pt}%
    \begin{overpic}[width=0.7\linewidth]{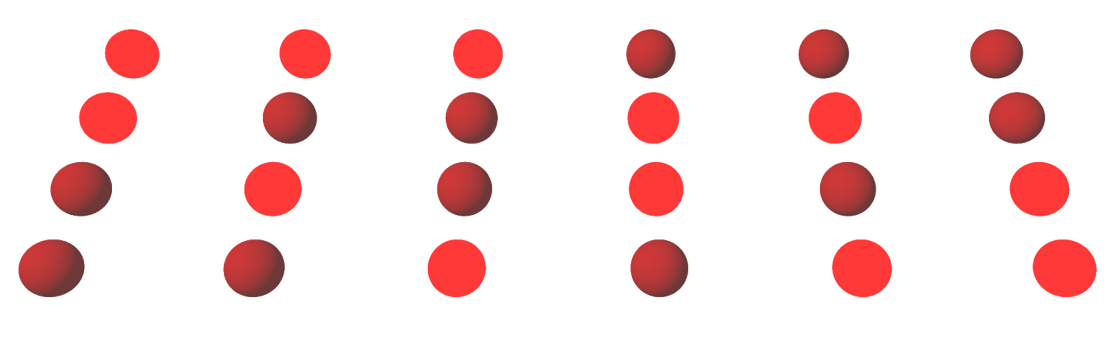}
    
    \put (-17,17) {\Large${\ket{\psi}=}$}
    \put (14.5,17) {\Large$-$}
    \put (31,17) {\Large$+$}
    \put (48,17) {\Large$-$}
    \put (65,17) {\Large$+$}
    \put (82,17) {\Large$-$}
    \end{overpic}
    \end{subfigure}
    \caption{(a) The negativity $\mathcal{N}$ of the LDBSS for increasing numbers of emitters when only weak pumping is applied to the superradiant emitters for even (thick) and odd (dashed) numbers of emitters. (b) The logarithmic negativity ($E_N=\log_2(2\mathcal{N}+1)$) of the LDBSS for varying relative strengths of $\gamma_P,\gamma_D,\gamma_I$. This is taken for a 1500-emitter system. (c) A graphic depiction of a maximally entangled four-emitter singlet state ($j=0$), where lighter spheres represent excited emitters and darker spheres non-excited.}
    \label{fig:EntangMeas}
\end{figure}

In Fig.~\ref{fig:EntangMeas}(a), we plot the negativity for increasing numbers of emitters, when $\gamma_D=\gamma_I=0$. We note a disparity between odd and even numbers of emitters, as the minimum total pseudo-spin states are $j=0$ and $j=1/2$ for even and odd numbers of emitters, respectively. Intriguingly, the negativity increases monotonically as the thermodynamic limit is approached; this is due to pumping ratcheting excitations up the dark $\kb{j,-j}$ states, moving to smaller values of $j$ and thus the fractional $z$-projection $j/N\rightarrow0$, increasing the negativity of such states. In Fig.~\ref{fig:EntangMeas}(b), we calculate the negativity for a 1500-emitter system at different relative rates for $\gamma_P,\gamma_D,\gamma_I$. We find that the negativity is maximal when pumping is the dominant effect, as it reduces $j$. Conversely, dephasing and independent decay effectively remove excitations from the system, increasing $j$.
 
 We can explore the thermodynamic limit for the pumping-only regime by inverting the relation from the top of the ladder, such that we consider the relation for $k = \lfloor{N}/{2} \rfloor-n$, and in the $N\rightarrow\infty$ limit, we can then formulate the recursion relation $f(k)$ and $g(k)$, independent of $N$ (see SM for expressions). Solving this recursion relation, one finds that for both odd and even numbers of emitters, more than $99.9\%$ of the population resides in the top five levels. As such, $\langle n\rangle \approx N/2$ in the large $N$ limit. For these states, we find that the negativity is $\mathcal{N}=0.235, 0.225$ for odd and even numbers of emitters, respectively. This is remarkably high, as the maximum attainable negativity across this bipartition is $\mathcal{N}=1/2$. Furthermore, this accumulation is toward multi-emitter singlet states $j=0$ that are relevant in field-gradient metrology~\cite{Urizar2013}; a depiction of these multi-emitter singlet states is given in Fig.~\ref{fig:EntangMeas}(c). Our quantitative entanglement result is consistent with numerical studies using entanglement witnesses~\cite{Shankar2021}. In Fig.~\ref{fig:EntangMeas}(a), we also plot the negativity of several maximal Dicke states $D_N^m$. We find that despite the LDBSSs being mixed states, they have comparable entanglement to these Dicke states.

The results presented here assume fully symmetric emitters, allowing a permutation-symmetric description. We expect asymmetry and disorder among the emitters to reduce the entanglement predicted by this model. Moreover, such disorder is likely to become more significant as the number of emitters increases. Further, the validity of the Lindblad description becomes questionable in the very high-number regime. Nevertheless, the key features of the entanglement structure identified here saturate for systems containing on the order of tens to hundreds of emitters. For weak disorder, its effects can be treated perturbatively and may be further mitigated by incorporating structured inter-emitter interactions.

In conclusion, we have examined the weak effects of symmetric local operations on superradiant emitters and how this breaks the degeneracy of the superradiant Liouvillian steady states. We have shown that these LDBSSs, besides the ground state, always possess a degree of entanglement, enabling the generation of entanglement from non-entangled states. As a corollary, we found that any permutation-invariant state with support outside the maximal Dicke ladder will have an entangled steady state. In the case of dominant pumping, these are highly entangled states with probability accumulating towards the multi-emitter singlet states $j=0$ that may be relevant for field-gradient metrology. We also established results in the thermodynamic limit, where we determined that these states follow a normal distribution with the same mean and variance as independent emitters. The non-Hermitian perturbation theory developed here is readily applicable to a broad class of related systems, including the single-mode Dicke model~\cite{Kirton2017} and multimode cavity-QED Dicke models recently employed to realize associative memory in quantum-optical spin glasses~\cite{Marsh2026}.

\acknowledgements{
The authors thank Julian Wiercinski, Moritz Cygorek and Peter Kirton for useful discussions. 
 The authors thank the Leverhulme Trust for support through grant number RPG-2022-335 and the Volkswagen Foundation - 0200195.
}

\bibliography{main}
\bibliographystyle{ieeetr}


\newpage
\widetext
\begin{center}
\textbf{\large  Entanglement generation from degeneracy breaking of superradiant Liouvillian steady states by local operations: Supplementary Material}
\end{center}
\setcounter{equation}{0}
\setcounter{figure}{0}
\setcounter{table}{0}
\setcounter{page}{1}
\makeatletter
\renewcommand{\theequation}{S\arabic{equation}}
\renewcommand{\thefigure}{S\arabic{figure}}
\renewcommand{\bibnumfmt}[1]{[S#1]}
\renewcommand{\citenumfont}[1]{S#1}
\section{Degenerate non-Hermitian perturbation theory}
To study the weak effects of the dynamical processes outlined above, we deploy degenerate non-Hermitian perturbation theory, where we assume $\gamma_{P,D,I}\ll\gamma_{o}$, and that the total Lindbladian is given by
\begin{equation}
    \mathcal{L} = \mathcal{L}_{o}+\epsilon(\mathcal{L}_D+\mathcal{L}_{P}+\mathcal{L}_{I}),
\end{equation}
to first order in $\epsilon$ the degeneracy of the superradiant steady states is lifted.
First-order perturbation theory for these processes breaks the degeneracy of the superradiant optical Liouvillian $\mathcal{L}_o$.
Consider a perturbative expansion of a steady state solution 
\begin{equation}
    \rho = \rho^{0}+\epsilon\rho_1 + O(\epsilon^2),
\end{equation}
clearly at zeroth order we have 
$\mathcal{L}_{o}\rho_0=0$,
which are the steady state solutions for the superradiant decay, which are the states $\kb{j,-j}\in \text{ker}(\mathcal{L}_{o})$. This is clearly a degenerate null space for all $j$ yield an element in the null space. To first order, we note that 
\begin{equation}
    \mathcal{L}_{o}\rho_1=-(\mathcal{L}_D+\mathcal{L}_{P}+\mathcal{L}_{I})\rho_0.
\end{equation}
Now as $\mathcal{L}_{o}$ is non-invertible we consider its left null vectors $Y_a$ that satisfy
\begin{equation}
    \mathcal{L}^\dag_{o}(Y_a) = 0.
\end{equation}
For a Lindblad master equation, the conjugate of the Lindbladian yields the Heisenberg equations and thus, $Y_a$ are the conserved quantities of the Lindblad master equation. For superradiant decay, these are $j$ values and, as such, the left null vectors are the projectors into fixed $j$ manifolds $Y_a=\sum_m \kb{j_a,m}$. It then follows that 
\begin{equation}
    \Tr\{Y_a^\dag \mathcal{L}_{o}\rho_{1}\}= 0  = -\Tr\{Y_a^\dag (\mathcal{L}_D+\mathcal{L}_{P}+\mathcal{L}_{I})\rho_{0}\}.
\end{equation}
The perturbation can be determined by finding the nullspace of the operator 
$A_e=Y(\mathcal{L}_D+\mathcal{L}_{P}+\mathcal{L}_{I})X$, where columns of $X$ are given by the elements of the kernel of $\mathcal{L}_{o}$ such that  $X_i = \kb{j_i,-j_i}$, $Y$ is determined by the elements of the leftnull space with $Y_i=\sum_m\kb{j_i,m}$ such that $A_e: \text{ker}\{\mathcal{L}_o\}\rightarrow \text{ker}\{\mathcal{L}_o\}$.  We can calculate the elements of this matrix as 
\begin{align}
    [A_e]_{j',j} &= \Tr\{\kb{j',-j'}A_e\kb{j,-j}\}\nonumber\\&=\delta_{j,j'}\tau_j+\delta_{j-1,j'}\beta_j + \delta_{j+1,j'}\kappa_j,
\end{align}
where $a_j =\sum_{i=P,D,I}a^{j,-j}_i$ for $a=\tau,\beta,\kappa$ with values~\cite{Chase2008}
\begin{align}
    \tau^{j,-j}&=\frac{1}{2j}(1+\frac{\alpha_{j+1,N}(2j+1)}{d_{j,N}(j+1)})A_{j,-j},\\
    \beta^{j,-j}&= \frac{\alpha_{j,N}}{2jd_{j,N}}B_{j,-j},\\
    \kappa^{j,-j} &=\frac{\alpha_{j+1,N}}{2(j+1)d_{j,N}}D_{j,-j},\\
    \alpha_{j,N} &= \frac{N!}{(N/2-j)!(N/2+j)!},\\
    d_{j,N} &= \frac{(2j+1)N!}{(N/2-j)!(N/2+j+1)!},\\
    A_{j,m} &= \gamma_P(j-m)(j+m+1)+4\gamma_D m^2,\\
    B_{j,m} &= \gamma_P(j-m)(j-m-1),\\
    D_{j,m} &= \gamma_P(j+m+1)(j+m+2)+4\gamma_D(j+m+1)(j-m+1)+\gamma_I(j-m+1)(j-m+2).
\end{align}
The thermodynamic limit scaling behaviour for the rates are 
\begin{align}
    \kappa^{j,\infty}&\approx   \gamma_I\frac{1}{2}(N-2{j}) ,\\
    \tau^{j,\infty}&\approx   N\gamma_D,\\
    \beta^{j,\infty}&\approx   \gamma_P\frac{1}{2}(N+2{j}).
\end{align}
We can compare this with the optical decay rate into the $\kb{j,-j}$ states $\Gamma^{j,-j}=2\gamma_oj$, thus, as $j\sim N/2 $, all rates scale linearly with $N$ and thus if they are perturbative at one value of $N$ they are perturbative for all.

Therefore, we consider the general solution 
\begin{equation}
    \rho_0 = \sum_j p_j\kb{j,-j},
\end{equation}
and we can solve for the null-space of $A_e$ by solving $A_e\rho_0=0$, this nullspace is then the first order degeneracy lifted steady states of the superradiant Lindbladdian. Applying $A_e$ to $\rho_0$ yields the equation 
\begin{align}
    A_e\rho_0 = \sum_j p_j (\tau_j\kb{j,-j} +\beta_j\kb{j-1,-j+1}+ \kappa_j\kb{j+1,-j-1})=0
\end{align}
 by relabelling the spin states with states that count the number of excited emitters  $\kb{n}=\kb{N/2-n,n-N/2}$,  
\begin{equation}
    \rho_0 = \sum_n p_n\kb{n},
\end{equation}
and comparing coefficients generates a recursion relation for the state probabilities in the simple form
\begin{equation}
    p_n = f_{n}p_{n-1}+g_np_{n-2}, \label{eqn:recursion}
\end{equation}
where substituting in the values for $\tau,\beta,\kappa$ yields

\begin{equation}
    f_n = \frac{(N-2n+2) (N-2n+1) \left({\gamma_P} \left(2 n^2-3 n (N+2)+N (N+4)+2\right)+((n-1) (4 {\gamma_D}+{\gamma_{I}} (N-2 n+4)))\right)}{n (N-2 n+4) (2 {\gamma_P}+(N-2n+1) (4 {\gamma_D}+{\gamma_{I}} (N-2 n+2)))}
\end{equation}
\begin{equation}
    g_n = \frac{-{\gamma_P} (N-n+3) (N-2n+3) (N-2n+2) (N-2n+1)}{n (N-2n+5) (2 {\gamma_P}+(N-2n+1) (4 {\gamma_{D}}+{\gamma_{I}} (N-2 n+2)))}.
\end{equation}

This recursion relation can then be numerically solved to find the steady state for varying strengths of the independent dynamical process rates by using seed values $p_{-1}=0, p_0=g$ and normalising the total probability. 

We can explore the thermodynamic limit for the pumping-only regime by inverting the relation from the top of the ladder, such that we consider the relation for $k = \lfloor\frac{N}{2} \rfloor-n$, and in the $N\rightarrow\infty$ limit,
We find the recursion coefficients take the form
\begin{align}
    f(k)= 
     \begin{cases}
        (1+k)^2(1+2k)(2+k)^{-1}, &k \text{ even}  \\
        (1+k)(3+2k)^2(5+2k)^{-1},&k\text{ odd}
    \end{cases}
\end{align}
\begin{align}
    g(k)= \begin{cases}
        -(1+k)(1+2k)(3+2k)(5+2k)^{-1}, &k \text{ even} \\
        -(1+k)(2+k)(3+2k)(3+k)^{-1},&k\text{ odd}
    \end{cases}.
\end{align}

\section{Entanglement at the Steady State}

As explored in Ref.~\cite{Rosario2025}, superradiance along the maximum total spin manifold (the Dicke ladder) alone does not guarantee entanglement generation, as the mixed states of the dynamics from the fully inverted state could be represented in a non-entangled coherent spin state basis. However, due to the mixing of the lowest $m=-j$ states, we expect that it is possible to generate entanglement in such cases.
 Furthermore, we wish to calculate the degree of entanglement of these systems; to this end, we utilise the negativity $\mathcal{N}$, an entanglement monotone, where $\mathcal{N}>0$ implies that the state is entangled. The negativity of a state $\rho$ is defined as 
 \begin{equation}
     \mathcal{N}(\rho) = \sum_i\frac{|\lambda_i|-\lambda_i}{2},
 \end{equation}
 where $\lambda_i$ are the eigenvalues of the density matrix $\rho^{\perp A}$, which is the partial transpose of the state $\rho$ with respect to a bipartition across $A$. 
 
 In particular, we will seek to see if any single emitter is entangled with the remaining $(N-1)$ emitter system. To this end, we note that using the Clebsch-Gordan coefficients
\begin{equation}
    \ket{j,-j} = \sum_{k,l=\pm 1/2}\sqrt\frac{d_{j+k}^{N-1}}{d_j^N}C_{j,1/2,k,l}\ket{1/2,l}\otimes\ket{j+k,-j-l}.
\end{equation}
This yields a density matrix in the site basis for the one emitter and the spin basis for the remaining $(N-1)$ emitters of the form

\begin{equation}
\rho = \sum_jp_j\sum_{i,k,l=\pm1/2}\frac{d_{j+i}^{N-1}}{d_j^N} C_{j,1/2,i,k}C^*_{j,1/2,i,l} \ket{1/2,k}\otimes\ket{j+i,-j-k} \bra{1/2,l}\otimes \bra{j+i,-j-l} 
\end{equation}
with Clebsch-Gordan coefficients 
\begin{equation}
    C_{J,1/2,i,j}= \begin{cases}
        -\sqrt{1/2(1+J/(J+1)}, & i=j=1/2\\
        \sqrt{1/2(1-J/(J+1)}, & -i=j=1/2 \\
        1, & i=j=-1/2 \\
        0, & \text{otherwise} \\
    \end{cases}
\end{equation}
performing the partial transpose yields
\begin{equation}
\rho^\perp = \sum_jp_j\sum_{i,k,l=\pm1/2}\frac{d_{j+i}^{N-1}}{d_j^N} C_{j,1/2,i,k}C^*_{j,1/2,i,l} \ket{1/2,k}\otimes\ket{j+i,-j-l} \bra{1/2,l}\otimes \bra{j+i,-j-k}, 
\end{equation}

From this, we need only calculate the eigenvalues of $\rho^\perp$, which can be calculated numerically. For individual states $\kb{j,-j}$ we have in the basis $\{\ket{1/2,1/2},\ket{1/2,-1/2}\}$ that 
\begin{align}
    \rho &= \begin{pmatrix}
        A &B\\C&D
    \end{pmatrix}\\
    A &= c_{j++}^2\ket{j_+,-j_+}\bra{j_+,-j_+},    \\
    B &= c_{j++}c_{j+-}\ket{j_+,-j_+}\bra{j_+,-j_-} ,\\
    C&=B^\dag,\\
    D &=  c_{j--}^2\ket{j_-,-j_-}\bra{j_-,-j_-}\\&+ c_{j+-}^2\ket{j_+,-j_-}\bra{j_+,-j_-},
\end{align}
where the normalised Clebsch-Gordan coefficients are $c_{jkl}= \sqrt\frac{d_{j+k}^{N-1}}{d_j^N}C_{j,1/2,k,l}$. Performing the partial transpose yields
\begin{equation}
    \rho^\perp = \begin{pmatrix}
          A &C\\B&D
    \end{pmatrix},
\end{equation}
finding the eigenvalues amounts to solving the characteristic polynomial $\det (\rho^\perp-\lambda \mathbf{1})=0$, which can be solved using Schur's complement formula as 
\begin{equation}
    \det(A-\lambda\mathbf{1})\det(D-\lambda\mathbf{1}-B(A-\lambda\mathbf{1})^{-1}C)=0,
\end{equation}
it is simple to see that the first term generates no such negative eigenvalues, and thus we concern ourselves only with the remaining term. Solving for $\lambda$ we can find that the only negative eigenvalue is 
\begin{equation}
    \lambda = c_{j++}c_{j+-} = -\frac{N-2j}{2N\sqrt{1+2j}}.
\end{equation}
thus it follows that the negativity of a single state is 
\begin{equation}
    \mathcal{N}(\kb{j,-j})=\frac{1}{2\sqrt{1+2j}}(1-\frac{2j}{N}).
\end{equation}
Generalising to the mixed states and noticing that subspaces become coupled, one can readily find, by the same procedure and much algebra,
\begin{equation}
    \mathcal{N}(\rho) = \sum_j \frac{1}{2}(\sqrt{c_{j--}^4p_j^2+4p_{j-1}^2c_{j-1+-}^2c_{j-1++}^2}-p_jc_{j--}^2)
\end{equation}
expanding in $j$-variables yields
\begin{equation}
    \mathcal{N}(\rho) = \sum_j \frac{1}{2(1 + 2 j) N}({\sqrt{\frac{(j^2 (1 - 2 j) (2 + 2 j + N)^2 p_j^2 + (1 + 2 j)^2 (2 - 2 j + 
    N)^2 p_{j-1}^2)}{1 - 2 j}}}-p_jj(2 + 2j + N).
\end{equation}

\section{Entanglement of $j=N/2$ Dicke states}
We can note that the multiplicity of the $j=N/2$ Dicke states is 1, yielding unique pure states. Also, we have that for the state 
\begin{equation}
    \ket{N/2,m}=\ket{D^{m=k-N/2}_N} =\frac{1}{\sqrt{\binom{N}{k}}} \sum_{\{k_i\}}\ket{k_i},
\end{equation}
where $\{\ket{k_i}\}$ are the set states with $k$ excitations. One can then rewrite this as 
\begin{equation}
    \ket{D^{m=k-N/2}_N} = \sqrt{\frac{k}{N}}\ket{1/2,1/2}\ket{D_{N-1}^{m-1}}+\sqrt{\frac{N-k}{k}}\ket{1/2,-1/2}\ket{D_{N-1}^m}. 
\end{equation}
From this, we can readily calculate the negativity of such a state as 
\begin{equation}
    \mathcal{N}(\ket{D_N^{m}} )=\frac{1}{2}\sqrt{1-\big(\frac{2M}{N}\big)^2}.
\end{equation}
\section{Thermodynamic Limit}

We now wish to explore the large-emitter regime by considering the quasi-continuous variable $x=n/N$ and taking the limit as $N\rightarrow \infty$. This leads to the relation 
\begin{align}
    f(x)  &= 1-g(x),\nonumber\\
    g(x) &= -\frac{(1-x)\hat{r}}{x}
\end{align}
$\hat{r}=\gamma_p/\gamma_o$. Interestingly, we note that in the large-$N$ limit, the effects of dephasing are completely suppressed, leaving only the pumping and independent decay terms. We now utilise a WKB approximation scheme by assuming that the probability has the following form in the thermodynamic limit 
\begin{equation}
    p_n=p(x) \propto e^{NS(x)}. 
\end{equation}
We can approximate the relative terms in the recurrence relation by considering the Taylor expansion to first order of $S(x)$ as 
\begin{equation}
    p_{n-k} = p(x-\frac{k}{N})\propto e^{NS(x)-kS'(x)}.
\end{equation}
The recurrence relation is now converted to a differential equation that takes the form 
\begin{align}
    e^{NS(x)}&= f(x) e^{NS(x)-S'(x)}+g(x)e^{NS(x)-2S'(x)}\\
    \implies  1&= f(x) e^{-S'(x)}+g(x)e^{-2S'(x)},
\end{align}
we can then solve for $S'(x)$ utilising the quadratic formula 
\begin{equation}
    S'(x) = -\ln\{ \frac{-f(x)+\sqrt{f^2(x)+4g(x)}}{2g(x)}  \}= \ln\{|g(x)|\}.
\end{equation}
From this, we can determine that the maximum in the probabilities will occur when $S'(x)=0$, which is exactly when $x=x'=\frac{\hat{r}}{1+\hat{r}}$. Using Laplace's method, we can determine that for $x'<0.5$ the probability distribution is given by
\begin{equation}
    p(x)\approx \left(\frac{2\pi x'^2}{N\hat{r}}\right)^{-1/2} \exp\{{-N\frac{\hat{r}}{2x'^2}\big(x-x'\big)^2}\}.\label{eqn:WKB}
\end{equation}

That is, the probabilities of the spin states follow a normal distribution with mean $\mu=x'$ and variance $\sigma^2 =\frac{x'^2}{N\hat{r}}$. Clearly, as we increase the number of emitters, this becomes a $\delta$ function about the point $x'$. The average number of excitations is then naturally $\langle n\rangle = Nx'$, and the variance in the number of excitations is $\Delta n^2=N\frac{x'^2}{r}$.
\section{Growth of Entanglement}
Using a similar approach, one can find that in the absence of independent decay, the population follows a behaviour approximately
\begin{equation}
    p(x) \propto \exp{4Nx\log N}.
\end{equation}
Utilising the convexity of the negativity, i.e. for $\rho=\sum_ip_i\rho_i$
\begin{equation}
    \mathcal{N}(\rho) \leq \sum_i p_i\mathcal{N}(\rho_i) =\mathcal{N}_\uparrow(\rho).
\end{equation}
We can approximate the entanglement by a quasicontinuum 
\begin{equation}
    \mathcal{N}_\uparrow(\rho)\approx \int dx p(x)\mathcal{N}(x),
\end{equation}
where $\mathcal{N}(x)=\mathcal{N}(j=(Nx-N/2))$, this enables an approximation of the upper bound as 
\begin{equation}
 \mathcal{N}_\uparrow(\rho)\approx \frac{1}{2} -\frac{1}{N}\left({\frac{9}{64 \log ^3(N)}-\frac{1}{8 \log ^2(N)}+\frac{1}{4 \log (N)}}\right)-\frac{15}{128 \log ^3(N)}+\frac{3}{32 \log ^2(N)}-\frac{1}{8 \log (N)},
\end{equation}
showing $N^{-1}$ behaviour for intermediate numbers of emitters and then a logarithmic convergence for large numbers of emitters. 

\end{document}